\documentstyle[epsf,12pt,here,psfig]{article}

\newcommand\pubnumber{SLAC-PUB-8283}
\newcommand\pubdate{Oct, 1999}
\def\Title#1{\begin{center} {\Large #1 } \end{center}}
\def\Author#1{\begin{center}{ \sc #1} \end{center}}
\def\Address#1{\begin{center}{ \it #1} \end{center}}

\def\doeack{\footnote{Work supported by the Department of Energy,
                     contract  DE-AC03-76SF00515.}}
\def\SLAC{Stanford Linear Accelerator Center\\
    Stanford University, Stanford, California 94309 USA}
\newcommand\pubblock{\rightline{\begin{tabular}{l} \pubnumber\\
         \pubdate  \end{tabular}}}

\def\stacksymbols #1#2#3#4{\def\theguybelow{#2}
    \def\vp{\lower#3pt}
    \def\sp{\baselineskip0pt\lineskip#4pt}
    \mathrel{\mathpalette\intermediary#1}}
\def\intermediary#1#2{\vp\vbox{\sp
     \everycr={}\tabskip0pt
     \halign{$\mathsurround0pt#1\hfil##\hfil$\crcr#2\crcr
              \theguybelow\crcr}}}

\def\Journal#1#2#3#4{{#1} {\bf #2}, #3 (#4)}

\def\PLB{{\em Phys. Lett.}  B}
\def\PRL{{\em Phys. Rev. Lett.}}
\def\PRD{{\em Phys. Rev.} D}

\def\PRep{{\em Phys. Rep.}}

\def\msb{{\bar{\ssstyle M \kern -1pt S}}}
\def\etal{{\it et al.}}

\begin{document}
\begin{titlepage}
\pubblock
\vfill \Title{Search for Free Fractional Electric Charge Elementary Particles}
\vfill \Author{V.~Halyo, P.~Kim, E.~R.~Lee, I.~T.~Lee, D.~Loomba,
and M.~L.~Perl\doeack} \Address{\SLAC} \vfill

\begin{abstract}
We have carried out a direct search in bulk matter for free fractional
electric charge elementary particles using the largest mass single sample
ever studied--- about $17.4$~mg of silicone oil.  The search used an
improved and highly automated Millikan oil drop technique.  No evidence for 
fractional charge particles was found.  The concentration of particles with
fractional
charge more than $0.16$~e (e being the magnitude of the electron charge)
from the nearest integer charge is less than $4.71\times10^{-22}$ particles
per nucleon with $95\%$ confidence. 
\end{abstract}
\medskip

\vfill
\end{titlepage}
Direct observation of free fractional charge elementary particles would
be an undisputed signature of physics beyond the Standard Model.  In this 
paper we present the results of an improved Millikan oil drop experiment
 designed to look for
such particles.  The apparatus made it possible to generate and measure the
charges of multiple columns of multiple drops simultaneously, each drop being 
$7.6-11.0$~$\mu$m in diameter.  This allowed us to have a large throughput of
$4.17\times10^7$ drops or about $17.4$ mg, of silicone oil.  

In the Standard Model there are no fractional charge color singlet
particles.  However such particles are expected in physics beyond the
Standard Model such as superstring theory.
In heterotic superstring models there can be either gauge coupling
unification with color singlet fractional charge particles or no 
fractional charge particles but also no unification. In fact all 
superstring models built to date have unification at the price of introducing 
fractional charge particles~\cite{chelliken}.
Other models for fractional charge particles are
outlined in~\cite{ME}.
There is no reason for fractional electric charge elementary
particles to be necessarily excluded.
Our motivation for these bulk matter searches is the possibility that 
these particles may have been produced in the early universe and some 
abundance remains today.

There are however no confirmed discoveries of free particles with fractional
 electric
charge.  Searches have been made using accelerators, cosmic rays~\cite{Lyons}
 and in bulk matter~\cite{Smithrev},~\cite{JonesRev}.  
Searches in bulk matter fall into two classes: those
that attempt to concentrate the fractional charge particles before the
search~\cite{Joyce} and those that directly search through all of
a bulk matter sample~\cite{Smith1},~\cite{NM}.  
Our preference is for direct bulk matter searches because it is frequently 
difficult to make a reliable estimate
of the efficiency of the pre-concentration method.
Our previous search using about $1$ mg of silicone oil
set an upper limit of less than $4.76\times10^{-21}$ particles
per nucleon~\cite{NM}.
The largest mass sample previously used in a 
direct bulk matter search was $4.9$~mg of niobium~\cite{Smith2} again 
with a negative result.

In our experiment, drops are ejected through a silicon
micromachined orifice and fall through air under the influence of 
gravity and an alternating vertical electric field. 
The drops are imaged by a digital charge coupled device (CCD) camera 
interfaced to a computer. The same computer is used to simultaneously
collect and  analyze the data and to monitor and control the experiment.

The drop generators we used consist of a glass fluid reservoir
tube with a  micromachined silicon orifice plate having  $7$--$10~\mu$m hole
diameter which is thermally welded to the end of the 
tube~\cite{patent},\cite{ELEE}.
A piezoelectric transducer disk made from lead zirconate titanate 
is attached to the lower portion of the tube.
The dropper is  filled with $5$ cS silicone oil. 
Silicone oil was chosen because it has low vapor pressure and the right 
viscosity to generate stable drops.
Drop ejection is initiated by an electrical pulse that causes the
piezoelectric transducer disk to contract radially on the glass, 
forcing a drop to form. The diameter of the drops can be varied
by a factor of two by adjusting the pulse height and duration,
using the method described in~\cite{ELEE}. Once the parameters are set,
the drop diameter remains constant to better than $1\%$. 

The drops are generated at $4$~Hz producing two columns separated 
by $300~\mu$m.
Once the drops are produced they fall into an electric field produced
by a parallel plate capacitor formed by an upper square ground plate 
of dimension
$10$~cm$\times10$~cm and a lower round high voltage plate $7.62$~cm 
in diameter.
The plates are placed horizontally $0.81$ cm apart with rectangular slits of 
dimension $1.27$~cm$ \times0.08$~cm to allow the passage of  
multiple columns of drops.  
The chamber and optical components are mounted on a vibrationally damped
optical table.
The electric field plates and the dropper are contained within two layers
of transparent polycarbonate  shielding since the drops are sensitive to 
convection due to their small radius.

The drops are backlit by red LEDs strobed at $10$~Hz with a $56~\mu$s
 pulse width. The light is diffused by a ground glass screen to
create uniform illumination.
A $135$~mm focal length lens $18$~cm away from the dropper focuses 
the image of the drops onto the CCD camera.
The camera is used to image the positions of the  falling drops. 
The active region of the CCD is $6.4$~mm $\times 4.8$~mm 
($736\times 242$~pixels) where the $6.4$~mm edge is chosen along  
the trajectory of the falling drops to maximize the number of position 
measurements and to avoid image distortion caused by camera interlacing.
The optical system has a magnification of $2.7$ so that the actual  field of
view of the falling drops is  $2.37$~mm vertically and $1.77$~mm horizontally.
A high speed video framegrabber captures images  from the CCD camera for
 computer analysis.
The drops have an average terminal velocity of $1.3$--$3.2$~mm/s 
depending on the drop radius
so that each drop is in the field of view for $8$--$11$ sequential images.
In order to ensure the   control of  the 
experiment and to help rule out fractional charge artifacts,
temperature, manometer pressure, vibration and  humidity are monitored. 

To find the positions of the drops in an image, 
the analysis program first applies a brightness level threshold  to isolate the
relevant pixels. 
The pixels which are above the threshold  are used to calculate an approximate
optical center (centroid)  for each drop.
A high accuracy calculation is then done by using the $20$ darkest 
pixels in a $10\times 10$~pixel  window around the approximate
centroid  to calculate a precise centroid, where each pixel is weighted 
by its intensity after subtracting the background value.
Once the centroid positions of each image have been measured, the sequence of 
centroids corresponding to the trajectory of each drop is extracted from 
the stream of data by a tracking  algorithm.
The core of the algorithm  examines several consecutive images and
considers all possible combinations of centroids.  The combination which form
physically consistent trajectories are grouped to form the initial 
trajectory of a drop. 
Once an initial trajectory has been found, it is possible to predict the 
position of that drop in future images.  
If a centroid is found in the predicted position, it is associated with the 
appropriate drop. 
When a drop left the field of view it is passed to the analysis code.
The software is capable of online operation at high rates, and is not 
currently a limiting factor in this experiment.
The search has three data sets; in chronological order Set I consisted of
 $1.4$~mg of $7.6~\mu$m average diameter drops, Set II consisted of 
$10.1$~mg of $10.4~\mu$m average diameter drops, and Set III consisted of
 $5.9$~mg of $9.4~\mu$m drops. These data sets with different drop diameters 
helped us to verify that we understood our charge measurement process.

To understand how drop charge  and  mass are measured consider
a drop falling under the influence  of gravity  in the presence 
 of a vertical  electric field that alternates between two discrete states,
 up and down .
Since the  drop falls in air, it reaches a terminal velocity.
The two equations that govern the motion of the drops are given by Stoke's law:
\begin{eqnarray}
	m g+E_{\downarrow}Q &=& 6 \pi \eta r v_{\downarrow}\nonumber \\
	m g-E_{\uparrow}Q &=& 6 \pi \eta r v_{\uparrow} \,
\end{eqnarray} 
where $m$ is the drop mass, $Q$ is the drop charge, $r$ is the drop radius,
$\eta$ is the viscosity of air and $v_{\downarrow}, v_{\uparrow}$
are the measured terminal velocities 
of the drops for the two directions of the electric field,
$E_{\downarrow}, E_{\uparrow}$.
We define $v_{e}$ and $v_{g}$ to be 

\begin{equation}
 v_{e} = {(v_{\downarrow}\;-\; v_{\uparrow}) \over 2}\;\;\;\;\;\;\;\; 
 v_{g} =  {(v_{\downarrow}\;+\; v_{\uparrow}) \over 2}\;\;\nonumber \,
\end{equation}

We know the mass of the drop since the density of silicone oil 
is  known ($\rho_{oil}=913.0~{kg/m^3}$) and we measure the radius using 

\begin{equation}
                r = 3\;\;
        \sqrt{{\eta \over {2\; g\; (\rho_{oil}-\rho_{air})}}}\;\;\sqrt{v_g}\,
\label{off}
\end{equation}

Using the measured velocities we calculate the charge of the drop in units
of the electron charge e 
\begin{eqnarray}
    q \equiv \frac{Q}{e}&=& C \;\; v_{e} \;\; \sqrt{v_{g}}\,
\label{deficharge}
\end{eqnarray}
where C is 
\begin{equation}
        C  = {{{18\; \pi }\over e }\;\;\;
        \sqrt{2\over {(\rho_{oil}-\rho_{air})\;g}}
        \;\;\;{1\over {E_{\downarrow}+E_{\uparrow}}}
        \;\;\;\eta^{3/2}}\;\;\;\; \,
\label{defiC}
\end{equation}
The charge of the drop is calculated  by finding 
the best fit to  the  sequence of centroid position measurements .
In addition a variety of different physical effects had to be corrected to
 achieve the best
 required charge accuracy.

Two rectangular  slits in the center of the 
electric field plates,  which allow passage of the drops, 
cause a spatial nonuniformity in the  electric field.
This nonuniformity combined with  the induced dipole on the drop produce
small changes in the apparent terminal velocity.
This dipole force monotonically reduce the velocity of the drops
as the drops fell.
By measuring the gradient in the electric field we were able to calculate that
 this effect was $3.15\%$ of $v_{g}$.

There is also an aerodynamic effect on the trajectory of the drops. 
The air in the vicinity of the columns of falling drops is dragged downwards
changing the apparent terminal velocity of the drops.
The resulting steady flow of the air causes 
the drops  to reach maximum velocity halfway between the plates
and then decelerate.
The magnitude of the effect was $1.46 \%$ of $v_{g}$.

Since we have done our measurement $0.2$~mm higher than halfway between the 
plates, the two effects acted on the drops simultaneously with opposite sign 
which led to a change of $1.69\%$ of $v_{g}$.
We chose the center of our operating region to be where the two effects  
maximally cancel each other.
In the analysis, these two phenomena are corrected simultaneously by 
fitting the velocities of the drops to a second order polynomial.

There is an additional interesting aerodynamic effect.
Since we had an imbalance in the number of positively and negatively 
charge drops,
there was a net motion of the drops, and hence of the air, which oscillated 
with the alternating  electric field. 
This caused a shift in the measured charge  on the order of $ 0.1$~e 
which was corrected.

After applying the above corrections we require that all
drops used in the data sample meet the measurement criteria listed in Table 
~\ref{table1}.
The first cut removes drops with charges higher than $4.5$ since the
 measurement 
accuracy decreases with charge. The second cut removes drops with less 
than $6$ centroids in order to have at least two charge measurements per drop.
 The third cut checks the consistency of the charge within a drop and the 
last cut checks for drops with high residual to eliminate tracking artifacts.
\begin{table}[here]
\begin{center}
\begin{tabular}{|cc|}\hline\hline
cuts&Percentage removed\\
\hline
 $|q|<4.5$&$3.056\%$\\
 $N>6$&$0.215\%$\\
 $\delta q<0.2$&$0.342\%$\\
 R$<8\sigma_{v}$&$0.0399\%$\\\hline\hline
\end{tabular}
\caption{The drops used in the data sample were subjected to cuts on 
the charge {\it{q}}   ($|q|<4.5$~e), on the number of centroids {\it{N}} 
 ($N>6$), consistency of charge measurements of one drop  $\delta q$ 
($\delta q<0.2$~e),
and the residuals {\it{R}} ($R<8 \sigma_v$), where $\sigma_{v}$ is the 
uncertainty in the velocity due to Brownian motion. Percentages removed 
by each cut are in order of application.
\label{table1}}
\end{center}
\end{table}

\begin{figure}
\centerline{\epsfxsize=4in\epsfbox[17 158 597 568]{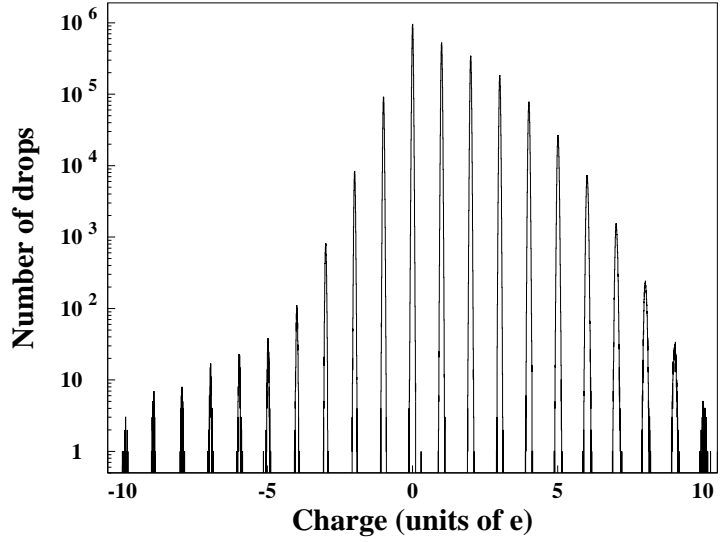}}
\caption{The charge distribution of $4.14\times10^7$ drops.}
\label{fig:q}
\end{figure}

These  criteria removed  $3.653\%$ of the total drops.
Figure~\ref{fig:q} shows the data after applying the last three cuts described
above, specifically $4.14\times10^7$ drops.
We see sharp peaks at integer numbers of charges and no drops with charges
further than $0.14$~e from the nearest integer charge, 
other than a single drop at $q=0.294$.

%%%%%%%%%%%%%%%%%%%%%%%%%%%%%%%%%%%%%%%%%%%%%%%%%%%%%%%%%%%%%%%%%%%%%%
%%%%%%%%%%%%%%%%%%%%%%%%%%%%%%%%%%%%%%%%%%%%%%%%%%%%%%%%%%%%%%%%%%%%%%

\begin{figure}
\centerline{\epsfxsize=4in\epsfbox[17 158 597 568]{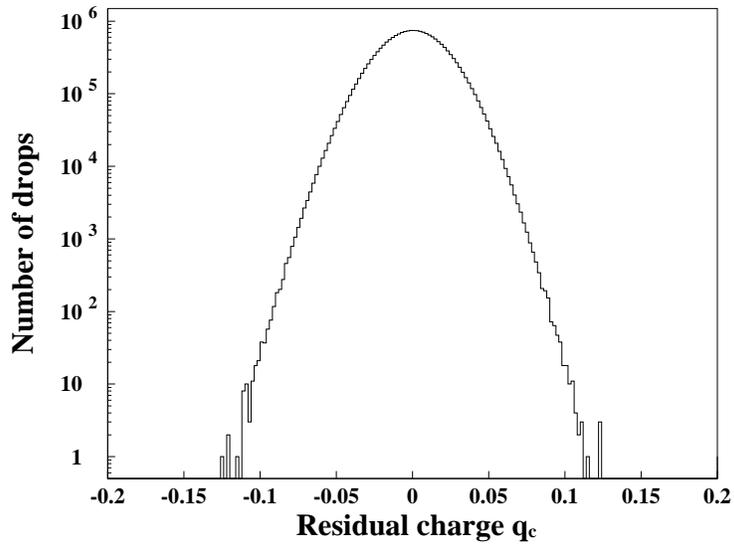}}
\caption{The residual charge based on  $1.885\times10^7$ drops.
	 The residual charge is defined as $q_{c}=q-N_{c}$.}
\label{fig:qc}
\end{figure}

Figure~\ref{fig:qc} shows the residual charge distribution of $q_{c}$,
which is defined as $q_{c}\equiv q-N_{c}$ where $N_{c}$ is the signed 
integer closest to  $q$, for data Set II.
It displays a superposition of integer charge peaks centered at zero.  
The peaks at each integer charge have a Gaussian distribution shape.
The standard deviation ($\sigma_{q}$) at charge zero is $0.018$e;
 higher charges result in a larger 
charge measurement error  since  $\sigma_{q}$  gets  contribution 
from terms involving ${v_{e}}/{v_{g}}$.
Table~\ref{table2}  lists the contributions to  $\sigma_{q}^2$;
the  contributions are from Brownian motion, centroid measurement and the
electric field non-uniformity between the plates.

%%%%%%%%%%%%%%%%%%%%%%%%%%%%%%%%%%%%%%%%%%%%%%%%%%%%%%%%%%%%%%%%%%%%%%
%%%%%%%%%%%%%%%%%%%%%%%%%%%%%%%%%%%%%%%%%%%%%%%%%%%%%%%%%%%%%%%%%%%%%%
\begin{table}[here]
\begin{center}
\begin{tabular}{|cccc|}\hline\hline
Source of Error& Set I&Set II&Set III\\
\hline
Brownian motion&$57.0\%$&$40.9\%$&$42.1\%$\\
Centroid measurement errors&$36.2\%$&$47.8\%$&$48.2\%$\\
Electric field non-uniformity&$6.8\%$&$11.4\%$&$9.7\%$ \\\hline\hline
\end{tabular}
\caption{The contribution to the charge measurement error $\sigma_{q}^2 $ for
each data.
\label{table2}}
\end{center}
\end{table}

\begin{figure}
\centerline{\epsfxsize=4in\epsfbox[17 158 597 568]{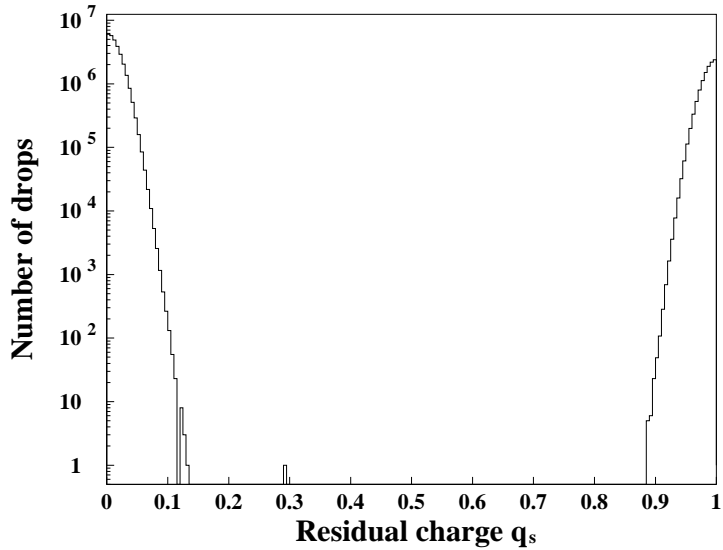}}
\caption{The residual charge on all the data. 
	 The residual charge  $q_{s}$ is defined $q_{s}=q-N_{s}$.}
\label{fig:qs}
\end{figure}

The search for drops with fractional charge is clarified 
in Figure~\ref{fig:qs} by the superposition of all data sets using 
the variable 
$q_s \equiv \left|q\right|-N_{s}$, where $N_{s}$ is defined to be the
 largest integer less than $\left|q\right|$.
This is the entire data remaining after the application of the cuts. 
There is no background subtraction.
Again one sees at $q_{s}=0.294$~e, the sole drop charge measurement that 
lies outside of the integer tails. We have applied the following experimental
 philosophy to this measurement.
In searching for a rare phenomenon it is important to apply the same
data selection criteria to all the data as we have done. The drop with 
 $q=0.294$~ fits all of our criteria and we do not know if it is the first 
indication for some background that begins to appear at the $1$ in
$4\times10^7$ level or if it has more significance.
Our only choice is to repeat the experiment with a larger sample and we 
intend to do so. 
%%%%%%%%%%%%%%%%%%%%%%%%%%%%%%%%%%%%%%%%%%%%%%%%%%%%%%%%%%%%%%%%%%%%%%
%%%%%%%%%%%%%%%%%%%%%%%%%%%%%%%%%%%%%%%%%%%%%%%%%%%%%%%%%%%%%%%%%%%%%%
%

\begin{figure}
\centerline{\epsfxsize=4in\epsfbox[17 158 597 568]{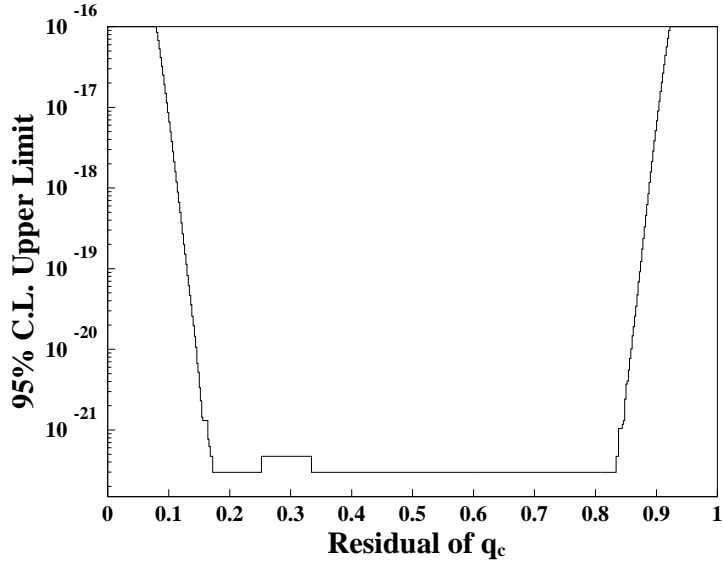}}
\caption{The $95\%$ C.L. upper limit on density of fractional charge
 particles per nucleon vs. residual of $q_{c}$.}
\label{fig:cl}
\end{figure}

Table~\ref{table3} presents  $95\%$ confidence upper limits on the number of
 fractional charge particles per nucleon in silicone oil for each data set.
We set conservative limits by counting the number of events in the signal
 region defined as within $2\sigma$ of each fractional charge,
 and calculating Poisson limits without background subtraction.
Figure~\ref{fig:cl} shows the combined $95\%$ confidence upper limits on the 
number of fractional charge particles per nucleon in silicon oil 
for the entire run.
We did not find any evidence for free fractional charge particles.  
We found with $95\%$ confidence that in
silicone oil the concentration of particles with fractional charge more than
$0.16$~e from the nearest integer charge is less than $4.71\times10^{-22}$ 
particles per nucleon except in the region  $0.26-0.34$~e where the upper limit
 is $2.98\times10^{-22}$ particles per nucleon.
\begin{table}
\begin{center}
\begin{tabular}{|ccccc|}\hline\hline
Data set &D ($\mu$m)&Mass (mg)&Range&Upper limit\\
\hline
I&$~7.6$&$1.35$&$0.15-0.26$&$3.84\times10^{-21}$\\
&&&$0.34-0.84$&$3.84\times10^{-21}$\\
&&&$0.26-0.34$&$6.07\times10^{-21}$ \\
II&$~10.4$&$10.13$&$0.16-0.84$&$5.12\times10^{-22}$ \\
III&$~9.4$&$5.92$&$0.17-0.86$&$8.76\times10^{-22}$ \\
&&&&\\
Total&&$~17.4$&$0.17-0.26$&$2.98\times10^{-22}$\\
&&&$0.34-0.84$&$2.98\times10^{-22}$\\
&&&$0.26-0.34$&$4.71\times10^{-22}$\\\hline\hline
\end{tabular}
\caption{Final result from the three runs of the experiment including the 
combined limit on the total mass examined, drop diameter D, sample mass, the
 range of residual $q_{c}$, the $95\%$ CL upper limit on the 
density of fractional charge particles per nucleon.}
\label{table3}
\end{center}
\end{table}

%%%%%%%%%%%%%%%%%%%%%%%%%%%%%%%%%%%%%%%%%%%%%%%%%%%%%%%%%%%%%%%%%%%%%%
%%%%%%%%%%%%%%%%%%%%%%%%%%%%%%%%%%%%%%%%%%%%%%%%%%%%%%%%%%%%%%%%%%%%%%
%

We have demonstrated several advantages of our Millikan method compared
to the levitometer method~\cite{Smith2} for searching for fractional charge
 particles in bulk matter.
The Millikan method allows a broad charge range to be studied with good charge
resolution and it provides natural self-calibration of the charge measurement.
It is amenable to automation and simple replication and it permits a 
relatively 
large amount of material to be examined. There is no obvious limit to the
 amount of material to be studied~\cite{dinesh}. 

Searches in bulk  refined matter such as silicone oil, niobium, or iron
 suffer from the uncertainty of whether a fractional charge particle would 
remain in the material during the chemical or physical refining 
process~\cite{ME},~\cite{Martinp}. Pure material also suffer from the
 uncertainty of whether the geochemical and geophysical
 processes that concentrate a mineral
in a local region of the Earth's crust would also carry along any  elementary
 fractional charge particles. Therefore, there is great value in
searching in unprocessed and unrefined bulk matter such as meteorites and
certain primordial terrestrial minerals. Our subsequent searches for 
fractional charge particles will use drops containing such materials.


\newpage
\begin{thebibliography}{99}

\bibitem{chelliken} A. N. Schellekens, \Journal{\PLB}{237}{363}{1990}.

\bibitem{ME} M. L. Perl and E. R. Lee, {\it Am. Journ. of Phys.} 
{\bf65},  698 (1997).

\bibitem{Lyons} L. Lyons, \Journal{\PRep}{129}{225}{1985}

\bibitem{Smithrev} P.F. Smith, {\it Ann. Rev. Nucl. Part. Sci.} {\bf 39}, 73-111 (1989).

\bibitem{JonesRev} L. Jones, {\it Rev. Mod. Phys.} {\bf 49}, 717 (1977).

\bibitem{Joyce} D. C. Joyce \etal, \Journal{\PRL}{51}{731}{1983}.

\bibitem{Smith1} P. F. Smith  \etal, \Journal{\PLB}{171}{129}{1986}.

\bibitem{NM} N. M. Mar \etal, \Journal{\PRD}{53}{11}{1986}.
Results from other major bulk matter searches are also listed.

\bibitem{Smith2} P. F. Smith \etal, \Journal{\PLB}{153}{188}{1985}.

\bibitem{patent} E. R. Lee and M. L. Perl, U.S. Patent 5943075 (1999). 

\bibitem{ELEE} E. R. Lee \etal, to be published.

\bibitem{dinesh} D. Loomba  \etal, submitted to {\it Rev. Scientific Instruments.}
\bibitem{Martinp} M. L. Perl,  \Journal{\PRD}{57}{7}{1988}.

\end{thebibliography}
\end{document}